\documentclass[a4paper,10pt,conference]{ieeeconf}
\IEEEoverridecommandlockouts 

\usepackage{cite,url}
\usepackage{graphicx,color}
\usepackage{amsmath,bm}
\newcommand\mymatrix[1]{\bm{\mathrm{#1}}}

\usepackage{tikz,pgflibraryshapes,pgflibraryarrows}

\begin{document}

\title{Analog Chaos-based Secure Communications and Cryptanalysis: \protect\\A Brief Survey%
\thanks{This paper has been published in the Proceedings of 3rd International IEEE Scientific
Conference on Physics and Control (PhysCon'2007, 3rd - 7th,
September 2007, Potsdam, Germany).}}
\author{Shujun~Li%
\thanks{Shujun Li was supported by the Alexander von Humboldt Foundation, Germany.}
\thanks{Shujun Li, Zhong Li and Wolfgang A. Halang are with
Fernuniversit\"at in Hagen, Chair of Computer Engineering, 58084
Hagen, Germany.},
Gonzalo~Alvarez\thanks{Gonzalo Alvarez is with Instituto de
F\'{\i}sica Aplicada, Consejo Superior de Investigaciones
Cient\'{\i}ficas, Serrano 144, 28006--Madrid, Spain.}, Zhong~Li
and Wolfgang~A.~Halang%
\thanks{Shujun Li is the corresponding author. Contact him via his personal web site: http://www.hooklee.com.}}

\maketitle
\thispagestyle{empty}

\begin{abstract}
A large number of analog chaos-based secure communication systems
have been proposed since the early 1990s exploiting the technique of
chaos synchronization. A brief survey of these chaos-based
cryptosystems and of related cryptanalytic results is given. Some
recently proposed countermeasures against known attacks are also
introduced.
\end{abstract}

\section{Introduction}

Since the late 1980s, chaos-based cryptography has attracted more
and more attention from researchers in many different areas. It has
been found that chaotic systems and cryptosystems share many similar
properties. For instance, chaotic systems are sensitive to the
initial conditions, which corresponds to the diffusion property of
good cryptosystems (for a comparison of chaos and cryptography, see
Table~1 in \cite{AlvarezLi:Rules:IJBC2006}). Basically, there are
two major types of chaos-based cryptosystems: analog chaos-based
secure communication systems and digital chaos-based ciphers, which
are designed employing completely different principles.

Almost all analog chaos-based secure communication systems are
designed based on the technique for chaos synchronization, which was
first discovered in the 1980s and then well developed in the 1990s
\cite{BKOVZ:ChaosSynchronization:PR2002}. The establishment of chaos
synchronization between two remote chaotic systems actually means
that some information has successfully been transmitted from one end
to the other. This fact naturally leads to the foundation of a
chaos-based communication system. Then, by keeping some part of the
involved chaotic systems secret, a third party not knowing the
secret key will not be able to reconstruct the information
transmitted. Thus, a chaos-based secure communication system is
created. Following this basic idea, a large number of analog
chaos-based secure communication systems have been proposed since
the 1990s. Meanwhile, related cryptanalytic work has also been
developed to evaluate performance (mainly the security) of various
analog chaos-based secure communication systems. Though a number of
surveys have been published to introduce progress in this area, they
become relatively obsolete due to the rapid growth of new research
work in recent years.

The purpose of this paper is to give a brief survey of analog
chaos-based secure communications and related cryptanalytic work,
especially focusing on latest work reported since the year 2000.
This paper is organized as follows. In the next section we first
introduce some preliminary knowledge about the underlying chaos
synchronization technique. Then, we classify most early chaos-based
secure communication systems into three basic types. Next, different
kinds of cryptanalysis are discussed with some concrete examples.
Finally, we enumerate some new countermeasures that have been
proposed to resist known attacks. A few concluding remarks are given
at the end of the paper to express our opinion on future trends in
this area.

\section{Chaos Synchronization}

Just as its name implies, synchronization of chaos denotes a process
in which two (or many) chaotic systems achieve a common dynamical
behavior after a period of transient period. Here, the common
behavior may be a complete coincidence of the chaotic trajectories,
or just a phase locking. To achieve synchronization, one or more
driving signals have to be sent from a source to the chaotic systems
to be synchronized. According to the source of the driving signal
and the mode of coupling, there are mainly four types of driving
modes:
\begin{itemize}
\item
directional (internal) driving: one chaotic system serves as the
source of driving and one or more driving signals are sent from this
systems to the others;

\item
bidirectional (internal) driving: two chaotic systems are coupled
with each other and are driven by each other in a mutual way;

\item
network-like coupling: many (more than two) chaotic systems are
coupled with others in some way to form a complex dynamic network;

\item
external driving: one or more external signals drive all the chaotic
systems involved towards a synchronized behavior.
\end{itemize}
Owing to the nature of secure communications (which means secret
information transmitted from one end to the other), directional
driving between two chaotic systems is employed for almost all
chaos-based secure communication systems. Therefore, in this section
we focus only on this kind of chaos synchronization.

For chaos synchronization of two chaotic systems with directional
driving, one of the chaotic systems serves as the master (or drive)
system, and the other is the slave (or response) system. From the
communication point of view, the master and slave system may also be
called sender and receiver system, respectively. For purpose of
chaos synchronization, one or more driving signals have to be
transmitted from the master system to the slave system as external
force to influence the the slave system's dynamics. As a result of
the driving force, the slave system may be able to follow the the
master system's dynamics exactly or in some other forms, thus
leading to different kinds of chaos synchronization like the
following ones that are widely used in chaos-based secure
communications:
\begin{itemize}
\item
complete synchronization (CS, also called identical
synchronization): the simplest form of chaos synchronization,
corresponding to a complete agreement of the trajectories of the
master and slave systems;

\item
generalized synchronization (GS): a generalized form of complete
synchronization for which the the slave system's trajectory
converges to the master's one in the sense of a one-to-one mapping
$f$;

\item
projective synchronization: a special case of GS with the one-to-one
mapping involved being a simple linear function
$f(\mymatrix{x})=\alpha \mymatrix{x}$;

\item
phase synchronization: the slave system matches its phase with that
of the master system, though their trajectories are not the same;

\item
lag synchronization: a time-delayed version of complete
synchronization for which the slave system coincides with the
time-delayed dynamics of the master system.
\end{itemize}
Due to other aspects of generating the driving signal, there are
also some other types of chaos synchronization. One of them is
called impulsive (or sporadic) synchronization, which means that the
driving signal is not transmitted to the slave system continuously,
but in an impulsive manner controlled by a fixed or time-varying
time interval $\tau$.

Another related concept called adaptive synchronization is a
technique that can help the slave system synchronize with the master
system in an adaptive way. This concept is useful not only for the
design of analog chaos-based secure communications, but also for the
cryptanalysis, because the adaption mechanism often implies that a
third-party can also drive its slave system to the sender and then
extract some secret information transmitted from the sender to the
legal receiver.

\section{Analog Chaos-based Secure Communications}

Most traditional analog chaos-based secure communication systems can
be classified into three basic types: chaotic masking, chaotic
switching (also called chaotic shift keying -- CSK) and chaotic
modulation. Although many new designs have been proposed in recent
years, most of them are actually modified or generalized
implementations of these three basic schemes. In this section, we
focus on the three basic schemes and give a brief summary of their
security. More details about related cryptanalytic results will be
the subject of the next section.

\subsection{Chaotic Masking}

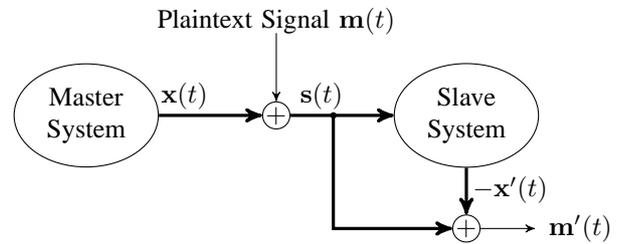
\begin{figure}[!htbp]
\centering
\begin{tikzpicture}[shape=ellipse,>=stealth']
\tikzstyle{every node}=[text badly centered];
\node [draw,text width=7ex] (Master) at (-2.5,0) {Master System};%
\node (Plaintext) at (0,1.25) {Plaintext Signal $\mymatrix{m}(t)$};%
\node [draw,circle,inner sep=0pt] (Plus1) at (0,0) {$+$};%
\node [draw,text width=7ex] (Slave) at (2.5,0) {Slave System};%
\node [draw,circle,inner sep=0pt] (Plus2) at (2.5,-1.5) {$+$};%
\node [inner sep=0pt] (Output) at (4,-1.5) {$\mymatrix{m}'(t)$};%
\draw[->,very thick] (Master) -- (Plus1) node[pos=0.25,above=-5pt] {$\mymatrix{x}(t)$};%
\draw[->,very thick] (Plus1) -- (Slave) node[pos=0.3,above=-5pt] {$\mymatrix{s}(t)$};%
\draw[->, shorten <=2pt] (Plaintext.base) -- (Plus1);%
\draw[->,very thick] (Slave.south) -- (Plus2) node[pos=0.9,above right,inner sep=2pt] {$-\mymatrix{x}'(t)$};%
\draw[->,very thick] (0.75,0) |- (Plus2);%
\draw [fill=black] (0.75,0) circle (1pt);%
\draw[->] (Plus2) -- (Output);%
\end{tikzpicture}
\caption{Basic structure of a typical chaotic masking
system.}\label{figure:ChaoticMasking}
\end{figure}

The earliest and simplest form of analog chaos-based secure
communication is chaotic masking, in which a plaintext message
signal $\mymatrix{m}(t)$ is embedded into a carrier signal
$\mymatrix{x}(t)$ to form a combined driving signal
$\mymatrix{s}(t)=\mymatrix{x}(t)+\mymatrix{m}(t)$, where the
addition operation ``$+$" can also be replaced by similar ones such
as multiplication. After chaos synchronization is established at the
receiver side, an estimation of $\mymatrix{x}(t)$ can be obtained
and, then, subtracted from $\mymatrix{s}(t)$ to recover the
plaintext signal. Figure~\ref{figure:ChaoticMasking} shows the basic
structure of a typical chaotic masking system.

To avoid the negative influence of the hidden plaintext signal on
chaos synchronization at the receiver side, the energy of the
plaintext message signal $\mymatrix{m}(t)$ should be much smaller
than that of the driving signal $\mymatrix{s}(t)$, i.e., much
smaller than the power of $\mymatrix{x}(t)$. Since the message
signal disturbs the driving signal, chaos synchronization cannot be
achieved exactly and, therefore, the message signal cannot be
recovered exactly. Another obvious feature of the chaotic masking
scheme is that the message signal does not influence the dynamics of
the master system at all.

The security of chaotic masking is questionable against various
attacks, mainly due to the fact that an attacker can always obtain
some information from the driving signal to construct (at least part
of) the dynamics of the master system. As the power energy of the
plaintext message must be much smaller than that of the driving
signal, it seems impossible to essentially eliminate this security
defect without changing the encryption structure.

\subsection{Chaotic Switching (Chaotic Shift Keying)}

\begin{figure}[!htbp]
\centering
\begin{tikzpicture}[>=stealth']
\tikzstyle{every node}=[text badly centered];
\node [draw,ellipse,text width=10ex] (Master0) at (-3,0.8) {Master System 0};%
\node [draw,rectangle,rounded corners,text width=10ex,inner sep=2ex] (Master1) at (-3,-0.8) {Master System 1};%
\node (Plaintext) at (0,2) {$\mymatrix{m}(t)=\{m(t)\in\{0,1\}\}_{t=0}$};%
\node [draw,ellipse,text width=8ex] (Slave) at (3,0) {Slave System};%
\node [draw,circle,inner sep=0pt] (Detector) at (3,-1.5) {$\times$};
\draw (0,0) circle(2pt) (-0.7,0.8) circle(2pt) (-0.7,-0.8) circle(2pt);%
\node [draw,circle,dotted,inner sep=0.85cm] (Switch) at (0,0) {};%
\draw[->,very thick] (2pt,0pt) -- (Slave) node[pos=0.6,below right,inner sep=1pt] {$\mymatrix{s}(t)$};%
\draw[very thick] (Master0) -- (-0.7cm-2pt,0.8cm) node[pos=0,above right,inner sep=1pt] {$\mymatrix{x}_0(t)$};%
\draw[very thick] (Master1) -- (-0.7cm-2pt,-0.8cm) node[pos=0.1,below right,inner sep=1pt] {$\mymatrix{x}_1(t)$};%
\draw[very thick] (-1.063,0) -- (-2pt,0pt);%
\draw[dashed,thick] (-2pt,0pt) -- node[right,inner sep=2pt] {$m(t)=0$} (-0.7cm+2pt,0.8cm);%
\draw[dashed,thick] (-2pt,0pt) -- node[right,inner sep=2pt] {$m(t)=1$} (-0.7cm+2pt,-0.8cm);%
\draw[->] (Plaintext) -- (Switch);%
\draw[<->] (130:0.6cm) arc(130:230:0.6cm);%
\draw[->,very thick] (Slave) -- (Detector) node[above right,inner sep=2pt] {$\mymatrix{x}'(t)$};%
\draw[->,very thick] (1.2,0) |- (Detector);%
\draw[fill=black] (1.2,0) circle(1pt);%
\draw[->] (Detector) -- (3,-2.5) node[left,inner sep=1pt]
{$\mymatrix{m}'(t)$};
\end{tikzpicture}
\caption{Basic structure of a typical chaotic switching (CSK) system
with $\otimes$ denoting the detector of chaos
synchronization.}\label{figure:CSK}
\end{figure}
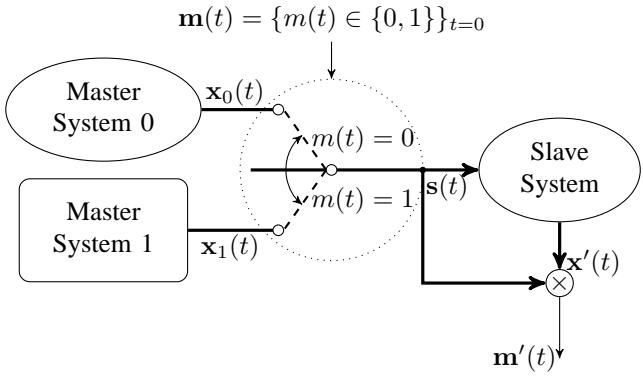

This scheme is mainly used to transmit digital signals. At the
sender side, two different chaotic systems are used for 0-bits and
1-bits of the plaintext message, respectively. That is, the employed
chaotic system is switched from time to time by the plaintext
message. At the receiver side, only one of the two chaotic systems
is needed, and the plaintext bits are recovered according to whether
or not the slave system can achieve chaos synchronization with the
master. Figure~\ref{figure:CSK} shows how a typical chaotic
switching system works to recover the plaintext message. Note that
the two chaotic systems at the sender side may be either homogeneous
or inhomogeneous. If two homogeneous ones are used, one chaotic
system with adjustable parameters suffices, which makes the
realization of chaotic switching systems more practical.

To ensure the establishment of chaos synchronization between the
master and slave systems, the transmission time of each plaintext
bit should be long enough. Therefore, the transmission rate of a
chaotic switching system is generally much slower than that of a
chaotic masking system. The main advantage of chaotic switching is
that the plaintext signal can exactly be recovered as long as the
level of the signal-to-noise ratio is not too low.

It has been known that the above simple chaotic switching system is
not secure against many different kinds of attacks. To enhance the
security, some modified chaotic switching systems have been proposed
in recent years, which will be discussed later in
Sec.~\ref{section:NewCountermeasures}.

\subsection{Chaotic Modulation}

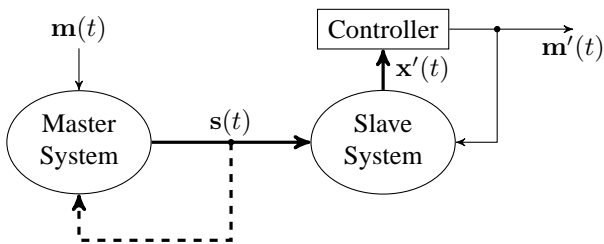
\begin{figure}[!htbp]
\centering
\begin{tikzpicture}[shape=ellipse,>=stealth']
\tikzstyle{every node}=[text badly centered];
\node [draw,text width=7ex] (Master) at (-2,0) {Master System};%
\node [inner sep=0pt] (Plaintext1) at (-2,1.5) {$\mymatrix{m}(t)$};%
\node [draw,rectangle,inner sep=4pt] (Controller) at (2,1.5) {Controller};%
\node [draw,text width=7ex] (Slave) at (2,0) {Slave System};%
\draw[->,very thick] (Master) -- node[above,inner sep=0pt] {$\mymatrix{s}(t)$} (Slave);%
\draw[->] (Plaintext1) -- (Master);%
\draw[->,very thick] (Slave) -- node[right,inner sep=0pt] {$\mymatrix{x}'(t)$} (Controller);%
\draw[fill=black] (0,0) circle(1pt) (3.5,1.5) circle(1pt);%
\draw[->,very thick,dashed] (0,0) |- (-2,-1.3) -- (Master.south);%
\draw[->] (Controller.east) -| (3.5,0) -- (Slave.east);%
\draw[->] (3.5,1.5) -- (4.5,1.5) node[below,inner sep=0pt]
{$\mymatrix{m}'(t)$};
\end{tikzpicture}
\caption{Basic structure of a typical chaotic modulation
system.}\label{figure:ChaoticModulation}
\end{figure}

Different from chaotic masking and chaotic switching schemes, in a
chaotic modulation scheme the plaintext message $\mymatrix{m}(t)$ is
injected into the sender system so that its dynamics is changed by
the plaintext message continuously. In this case, generally an
adaptive controller (which can also be considered as an extra
dynamical system bidirectionally coupled with the sender system) is
added at the slave system according to some rule such that its
output $\mymatrix{m}'(t)$ asymptomatically converges to
$\mymatrix{m}(t)$. To follow the master system's dynamics, generally
the controller's output (i.e., $\mymatrix{m}(t)$) should be injected
into the slave system in the same way as in the master. See
Fig.~\ref{figure:ChaoticModulation} for the basic structure of a
typical chaotic modulation system. Note that in some chaotic
modulation systems there may be no feedback of $\mymatrix{s}(t)$
back into the master system.

There are two different types of chaotic modulation: (1) parameter
modulation, in which the plaintext message signal $\mymatrix{m}(t)$
modulates the values of one or more control parameters; (2) direct
modulation, in which $\mymatrix{m}(t)$ is injected into one or more
variables of the master system without changing the value of any
control parameter. In some chaotic modulation schemes, the plaintext
signal is also embedded into the driving signal, which can be
regarded as a modified version of chaotic masking (via feedback of
the driving signal and some other necessary modifications).

Compared with chaotic masking schemes, chaotic modulation schemes
can exactly recover the plaintext signal (in an asymptotical manner)
if some conditions are satisfied. Considering that chaotic switching
systems can only transmit digital signals, chaotic modulation also
has a better performance than chaotic switching. In fact, carefully
designed, the chaotic modulation technique can even be used to
transmit more than one plaintext message signal. One possible way
for this is to modulate $n$ control parameters of the master system
with $n$ plaintext message signals, respectively.

The main disadvantage of chaotic modulation is that the controller
depends on the master and slave systems' structure, which means that
different controllers needs to be designed for different master
systems. Controllers may not even exist in certain cases for
essential defects of the master/slave chaotic systems.

\section{Cryptanalytic Results}

Many chaos-based secure communication systems were proposed without
much security analysis. The security of these cryptosystems was
simply ``ensured'' by the underlying chaotic systems' complexity.
From a cryptographer's point of view, however, the complexity of
chaos does not necessarily mean that a chaos-based cryptosystem is
secure. To evaluate the security of a cryptosystem, all known
cryptanalytic methods (i.e., attacks) have to be investigated
specifically for the target cryptosystem.

As a basic rule in cryptology, it is always assumed that all details
about the target encryption algorithm are known to the attacker
\cite[p. 5]{Schneier:AppliedCryptography96}. The secret key should
be the only component unknown to the attacker and used to guarantee
the cryptosystem's security.

The first step of cryptanalysis is to estimate the size of the key
space, i.e., to see if the complexity of exhaustively searching all
possible keys of a cryptosystem is not cryptographically high.
According to the computational power of today's computers, a key
space of size $O(2^{100})$ is generally required.

In the case that the key space is large enough, one needs to further
investigate the security of the cryptosystem against all known
attacks, which include the following four kinds of attacks
(classified according to the resources that an attacker can access):
\begin{itemize}
\item
\textit{ciphertext-only attack}: only the ciphertexts can be
observed by the attacker;

\item
\textit{known-plaintext attack}: some plaintexts and the
corresponding ciphertexts can be observed by the attacker;

\item
\textit{chosen-plaintext attack}: some plaintexts can be freely
chosen by the attacker and the corresponding ciphertexts can be
observed;

\item
\textit{chosen-ciphertext attack}: some ciphertexts can be freely
chosen by the attacker and the corresponding plaintexts can be
observed.
\end{itemize}
The chosen-ciphertext attack generally works only when the attacker
has temporary access to a legal decipher (receiver), which is a
mirror version of the chosen-plaintext attack at the encipher
(sender) side.

Since the mid-1990s, a large number of cryptanalytic results have
been reported on chaos-based secure communications. It has been
shown that most traditional schemes are not sufficiently secure from
a cryptographical point of view. In this section, we classify these
cryptanalytic results into several different categories.

\subsection{Low Sensitivity to Secret Key}

The most common (and maybe also the most serious) problem about
chaos-based secure communications is the low sensitivity to the
secret key (i.e., the control parameters of the master chaotic
system). The low sensitivity is a necessary requirement for real
implementations of any analog chaos-based cryptosystem, because it
is impossible to ensure exact matching of the master and slave
systems. Unavoidable noise and manufactural component deviation
involved in chaotic circuits are the two main factors causing this
security problem. According to recent results reported in
\cite{Wang:EFA:Chaos04, Zhang:EFA:Chaos06}, it has been verified
that most analog chaos-based secure communication systems suffer
from this defect.

As a direct result of this low-sensitivity problem, the size of the
key space becomes much smaller than expected. Therefore, a
brute-force attack can be mounted to approximately guess the secret
key, and the estimated key can be used later to approximately
decrypt the plaintext message signals.

\subsection{Parameter Estimation}

For most analog chaos-based secure communication systems, the low
sensitivity to the secret key is actually caused by a very simple
relationship between synchronization error and key mismatch: the
larger the key mismatch is, the larger the synchronization error
will be, and vice versa. This means that an iterative algorithm can
be used to determine the value of the secret parameters, which
corresponds to the concept of ``adaptive synchronization''. A lot of
work has been reported about adaptive synchronization when the
master system's parameters are unknown to the receiver. Some of the
work can directly be used or easily extended to break analog
chaos-based secure communication systems
\cite{Dedieu:PatameterIdentification:IEEETCASI97,
TaoDu:ChaoticCrytpanalysis:IJBC2003a,
TaoDu:ChaoticCrytpanalysis:IJBC2003b}.

Besides the method based on adaptive synchronization, there are also
other ways one can use to estimate the secret parameters (i.e., the
key) of the chaos-based cryptosystems. For instance, due to the
nature of Lorenz and Chua chaotic systems, the secret parameters can
be determined from the driving signal and its derivative (mainly
differentials of different orders)
\cite{Beth:ChaoticCryptanalysis:EuroCrypt94,
Vaidya:DecodingChaoticSuperkey:CSF2003,
Wu:ChuaParameterEstimation:PLA2004}. For some specific schemes, it
is also possible to derive part of the secret parameters by
analyzing the return maps of the master systems
\cite{LiAlvarez:CSF2005}.

When chosen-ciphertext attacks are possible, i.e., when the attacker
can access a legal receiver for some time, the attacker can set the
driving signal to a fixed constant $C$ in order to obtain the values
of all secret parameters \cite{Hu:CCA:IEEETCASI2003}.

\subsection{Estimating Carrier Signal}

When the plaintext message signal $\mymatrix{m}(t)$ is hidden in the
driving signal $\mymatrix{s}(t)=\mymatrix{x}(t)+\mymatrix{m}(t)$, it
may be possible to recover the approximate dynamics of the master
system and, then, obtain an estimation of the carrier signal
$\mymatrix{x}(t)$, thus leading to an approximate recovery of
$\mymatrix{m}(t)$. This idea works for chaotic masking and some
chaotic modulation systems.

The first report about this cryptanalytic method was proposed by
Short et al. in \cite{Short:UnmaskingChaos:IJBC94}, by employing the
NLD (nonlinear dynamics) forecasting technique to estimate the
master system's dynamics from the driving signal $\mymatrix{s}(t)$
of chaotic masking systems. Later he refined this technique and
extended it for chaotic modulation systems
\cite{Short:UnmaskingChaos:IJBC96,
Short:ChaoticSignalExtraction:IJBC97}. The NLD technique has been
widely employed to break many simple chaos-based secure
communication systems.

\subsection{Direct Extraction of Plaintext}

For some chaos-based secure communication schemes, it is also
possible to directly estimate the plaintext message signals from the
driving signals without estimating the secret key or the carrier
signals. Many different methods have been reported in recent years,
mostly for chaotic masking and chaotic switching schemes. In this
subsection, we introduce some of these specific methods.

\subsubsection{Return-map Analysis}

By constructing some return maps of the master system, one may be
able to estimate the plaintext message signal from the fluctuation
(for chaotic masking) and the splitting (for chaotic switching) of
the return maps. This method was first proposed in
\cite{Perez:ReturnMapCryptanalysis:PRL95} and further developed in
\cite{LiAlvarez:CSF2005, LiShujun:ReturnMapAttack:IJBC2006} for
other more advanced systems.

\subsubsection{Power-spectral (Filtering) Analysis}
\label{section:SpectrumCryptanalysis}

Though the dynamics of chaotic systems are rather complex, the power
spectra of their variables are not so complex as expected. As
investigated in \cite{LG:ModdingSymmetries:JPIIF96,
AlvarezLi:LorenzFrequency2004}, even when the power spectra of some
chaotic systems seem to be good, significant spectrum peaks can
still be founded in the spectra by removing the symmetries of the
chaotic attractors. For instance, the spectrum of $x(t)$ in the
Lorenz system is relatively good, but that of $|x(t)|$ has a
significant peak. When the plaintext message signal is hidden in the
driving signal, the narrow-band spectrum means that the driving
signal can be directly filtered to recover the message signal
\cite{Yang:SpectralCryptanalysis:PLA98,
AlvarezLi:BreakingPS:CSF2005}.

\subsubsection{Power Energy Analysis}

For some parameter modulation systems, the power energy of the
driving signal varies according to the value of the transmitted
signal. This makes it possible to obtain a smoothed version of the
message signal by observing the average power energy of the driving
signal in a sliding time-window \cite{Alvarez:BreakingCPM:CSF2004}.
Exact recovery of the plain message signal is possible for chaotic
switching systems, because each bit has to be held for some time to
ensure that chaos synchronization is established.

\subsubsection{Generalized Synchronization-based Method}

For chaotic switching systems and some parameter modulation systems,
there is a simple relationship between the synchronization error and
the value of the transmitted signal. This fact can be exploited to
extract the plaintext message signal directly
\cite{Yang:GSCryptanalysis:IEEETCASI97,
AlvarezLi:BreakingPS:CSF2005}.

\subsubsection{Short-time Period analysis}

When the spectrum of the driving signal (or its derivative of some
form) involved has a significant peak (recall
Section~\ref{section:SpectrumCryptanalysis}), generally there exists
a simple relationship between the peak frequency and the values of
the control parameters (see Fig.~9 in
\cite{AlvarezLi:LorenzFrequency2004}). In this case, one can try to
extract the short-time period as a measurement of the peak frequency
modulated by the plaintext message signal. According to the change
of the extracted short-time periods, the plaintext message signal
can be extracted exactly (for chaotic switching systems) or
approximately (for some parameter modulation systems)
\cite{Yang:STZCR:IJCTA95, AlvarezLi:ShortTimePeriod}.

\subsubsection{Switching-event Detection}

For chaotic switching and parameter modulation systems, the dynamics
of the master systems will change significantly when the value of
the modulating signal (i.e., the plaintext message signal) changes.
By detecting and tracking these switching events, it may be possible
to recover the modulating signal
\cite{Storm:DetectingSwitchingEvent:PRE2002}.

\section{New Countermeasures Against Known Attacks}
\label{section:NewCountermeasures}

To overcome the security problems of most traditional chaos-based
secure communication schemes, a number of new countermeasures have
been proposed in recent years. Among all the known attacks, the ones
based on and the NLD forecasting technique have received most
attention, while little work has done on the low sensitivity to the
secret key and the security problem of parameter estimation. This
section lists some of these new countermeasures and known
cryptanalytic results.

\subsection{Using More Complex Chaotic Systems}

One widely suggested measure is to use more complex chaotic systems
rather than three-dimensional systems like the Lorenz and Chua
systems. Hyperchaotic systems and time-delay chaotic systems have
been adopted for some newly designed chaos-based secure
communication systems. Unfortunately, a number of recent
cryptanalytic results have shown that the introduction of hyperchaos
or time-delay chaos cannot essentially enhance security
\cite{Short:UnmaskingHyperchaos:PRE98,
ZhouLai:ChaoticCryptanalysis:PRE99b,
Huang:UnmaskingChaosWavlet:IJBC2001,
TaoDu:ChaoticCrytpanalysis:IJBC2003a, AlvarezLi:BreakingPS:CSF2005}.

\subsection{Using More Complicated Synchronization Modes}

Another widely suggested measure is to use more complicated
synchronization modes, such as impulsive, projective, phase, or lag
synchronization, and so on. Although definitive results have not
been obtained for the overall performance of each new
synchronization mode, some security problems have been reported for
specific schemes \cite{Alvarez:BreakingPS:Chaos2004,
AlvarezLi:BreakingPS:CSF2005}. It seems that impulsive
synchronization is most promising as candidate base of designing new
chaos-based secure communication systems.

\subsection{Additional Encryption Functions}

In \cite{Yang:3CC:IEEETCASI97}, Yang et al. suggested adding an
additional encryption function (actually the $n$-fold composition of
a piecewise linear mapping) to chaos-based secure communication
systems to enhance the security. This idea was later employed by
some other researchers. Although there is not too much cryptanalytic
work about this kind of combined chaos-based secure communication
systems, a recent result about NLD
\cite{Short:ChaoticCrptanalysis:IEEETCASI2001} implies that the
additional encryption function may be circumvented.

\subsection{Combining Heterogeneous Chaos-based Cryptosystems}

By combining different types of chaos-based cryptosystems, the
security of the resulting system may be higher than the security of
each constituent. The simplest way of combination is to cascade two
or more heterogeneous chaos-based cryptosystems together, e.g., a
chaotic masking subsystem plus a chaotic modulation subsystem as
proposed in \cite{Murali:3CC:PhysicaD2000}, or a chaotic switching
subsystem plus a chaotic modulation subsystem as proposed in
\cite{Murali:CHCS:PRE2001}. Unfortunately, this simple combination
has been proved to be insecure
\cite{TaoDu:ChaoticCrytpanalysis:IJBC2003a}. So, more complicated
approaches of combination should be further investigated.

\subsection{Two-channel Approach}

In \cite{Jiang:2Channels:IEEETCASI2002}, two separate channels were
suggested to enhance the security: one channel only for chaos
synchronization and the other one for complicated encryption of the
plaintext message signal. This scheme has been found insecure
\cite{AlvarezLi:Breaking2Channels2006}, because parameter estimation
is still possible by analyzing the chaos synchronization channel.

\subsection{Remodulating the Driving Signal}

In \cite{BuWang:CSF2004} a countermeasure was proposed in form of
remodulating the driving signal before sending it to the receiver
side. This approach was soon broken \cite{CheeXuBishop:CSF2004,
WuHuZhang:CSF2004, Alvarez:BreakingCPM:CSF2005}, however, and one
improved version \cite{WuHuZhang:CSF2004} has also proved to be
insecure \cite{LiAlvarez:CSF2005} as well.

\subsection{Modified Chaotic Switching Schemes}

To enhance security against the return map attack, it is possible to
extend a chaotic switching system to include $2n>2$ chaotic systems
\cite{Indian:MSCPM:IJBC2001}, in which $n$ systems correspond to the
plaintext bit 0 and the other $n$ systems to the plaintext bit 1.
For each plaintext bit, the sender randomly chooses a system from
the $n$ candidates and the receiver checks all $2n$ chaotic systems
to find out the correct one. In \cite{Indian:MSCPM:IJBC2001} another
measure is also adopted to further enhance the security, viz., to
frequently change the driving signal from one variable to another. A
recent cryptanalytic report has shown that both countermeasures are
still not secure against the return map attack
\cite{LiShujun:ReturnMapAttack:IJBC2006}.

In \cite{XuChee:CSKwFS:IJBC2004}, pseudo-random false switching
events are introduced to enhance security against various known
attacks. No cryptanalytic result has been reported on this
countermeasure so far.

\section{Concluding Remarks}

As most traditional chaos-based secure communication systems and
many new-generation ones have been known to be insecure, novel ideas
need to be created to improve security. Combining more than two
countermeasures may be a promising way to get more secure
cryptosystem. The low sensitivity to the secret key and the
potential possibility to mount attacks based on parameter estimation
are regarded as the two greatest problems in almost all analog
chaos-based secure communication systems, thus deserving more
attention in future research.

\bibliographystyle{IEEEtran}
\bibliography{ref}

\begin{thebibliography}{10}
\providecommand{\url}[1]{#1}
\csname url@rmstyle\endcsname
\providecommand{\newblock}{\relax}
\providecommand{\bibinfo}[2]{#2}
\providecommand\BIBentrySTDinterwordspacing{\spaceskip=0pt\relax}
\providecommand\BIBentryALTinterwordstretchfactor{4}
\providecommand\BIBentryALTinterwordspacing{\spaceskip=\fontdimen2\font plus
\BIBentryALTinterwordstretchfactor\fontdimen3\font minus
  \fontdimen4\font\relax}
\providecommand\BIBforeignlanguage[2]{{%
\expandafter\ifx\csname l@#1\endcsname\relax
\typeout{** WARNING: IEEEtran.bst: No hyphenation pattern has been}%
\typeout{** loaded for the language `#1'. Using the pattern for}%
\typeout{** the default language instead.}%
\else
\language=\csname l@#1\endcsname
\fi
#2}}

\bibitem{AlvarezLi:Rules:IJBC2006}
G.~Alvarez and S.~Li, ``Some basic cryptographic requirements for chaos-based
  cryptosystems,'' \emph{Int. J. Bifurcation and Chaos}, vol.~16, no.~8, pp.
  2129--2151, 2006.

\bibitem{BKOVZ:ChaosSynchronization:PR2002}
S.~Boccaletti, J.~Kurths, G.~Osipov, D.~Valladares, and C.~Zhou, ``The
  synchronization of chaotic systems,'' \emph{Phys. Rep.}, vol. 366, no. 1-2,
  pp. 1--101, 2002.

\bibitem{Schneier:AppliedCryptography96}
B.~Schneier, \emph{Applied Cryptography -- Protocols, algorithms, and souce
  code in C}, 2nd~ed.\hskip 1em plus 0.5em minus 0.4em\relax New York: John
  Wiley \& Sons, Inc., 1996.

\bibitem{Wang:EFA:Chaos04}
X.~Wang, M.~Zhan, C.-H. Lai, and G.~Hu, ``Error function attack of chaos
  synchronization based encryption schemes,'' \emph{Chaos}, vol.~14, no.~1, pp.
  128--137, 2004.

\bibitem{Zhang:EFA:Chaos06}
Y.~Zhang, C.~Tao, and J.~J. Jiang, ``Theoretical and experimental studies of
  parameter estimation based on chaos feedback synchronization,'' \emph{Chaos},
  vol.~16, no.~4, p. art. no. 043122, 2006.

\bibitem{Dedieu:PatameterIdentification:IEEETCASI97}
H.~Dedieu and M.~J. Ogorza{\l}ek, ``Identifiability and identification of
  chaotic systems based on adaptive synchronization,'' \emph{IEEE Trans.
  Circuits and Systems I}, vol.~44, no.~10, pp. 948--962, 1997.

\bibitem{TaoDu:ChaoticCrytpanalysis:IJBC2003a}
C.~Tao, G.~Du, and Y.~Zhang, ``Decoding digital information from the cascaded
  heterogeneous chaotic systems,'' \emph{Int. J. Bifurcation and Chaos},
  vol.~13, no.~6, pp. 1599--1608, 2003.

\bibitem{TaoDu:ChaoticCrytpanalysis:IJBC2003b}
C.~Tao and G.~Du, ``A new approach to breaking down chaotic secure
  communication,'' \emph{Int. J. Bifurcation and Chaos}, vol.~13, no.~9, pp.
  2689--2698, 2003.

\bibitem{Beth:ChaoticCryptanalysis:EuroCrypt94}
T.~Beth, D.~E. Lazic, and A.~Mathias, ``Cryptanalysis of cryptosystems based on
  remote chaos replication,'' in \emph{Advances in Cryptology -
  {E}uro{C}rypt'94}, ser. Lecture Notes in Computer Science, vol. 950.\hskip
  1em plus 0.5em minus 0.4em\relax Spinger-Verlag, Berlin, 1994, pp. 318--331.

\bibitem{Vaidya:DecodingChaoticSuperkey:CSF2003}
P.~G. Vaidya and S.~Angadi, ``Decoding chaotic cryptography without access to
  the superkey,'' \emph{Chaos, Solitons and Fractals}, vol.~17, no. 2-3, pp.
  379--386, 2003.

\bibitem{Wu:ChuaParameterEstimation:PLA2004}
L.~Liu, X.~Wu, and H.~Hu, ``Estimating system parameters of {Chua}'s circuit
  from synchronizing signal,'' \emph{Physics Letters A}, vol. 324, no.~1, pp.
  36--41, 2004.

\bibitem{LiAlvarez:CSF2005}
S.~Li, G.~Alvarez, and G.~Chen, ``Breaking a chaos-based secure communication
  scheme designed by an improved modulation method,'' \emph{Chaos, Solitons \&
  Fractals}, vol.~25, no.~1, pp. 109--120, 2005.

\bibitem{Hu:CCA:IEEETCASI2003}
G.~Hu, Z.~Feng, and R.~Meng, ``Chosen ciphertext attack on chaos communication
  based on chaotic synchronization,'' \emph{IEEE Trans. Circuits and Systems
  I}, vol.~50, no.~2, pp. 275--279, 2003.

\bibitem{Short:UnmaskingChaos:IJBC94}
K.~M. Short, ``Steps toward unmasking secure communications,'' \emph{Int. J.
  Bifurcation and Chaos}, vol.~4, no.~4, pp. 959--977, 1994.

\bibitem{Short:UnmaskingChaos:IJBC96}
------, ``Unmasking a modulated chaotic communications scheme,'' \emph{Int. J.
  Bifurcation and Chaos}, vol.~6, no.~2, pp. 367--375, 1996.

\bibitem{Short:ChaoticSignalExtraction:IJBC97}
------, ``Signal extraction from chaotic communications,'' \emph{Int. J.
  Bifurcation and Chaos}, vol.~7, no.~7, pp. 1579--1597, 1997.

\bibitem{Perez:ReturnMapCryptanalysis:PRL95}
G.~P\'{e}rez and H.~A. Cerdeira, ``Extracting messages masked by chaos,''
  \emph{Physical Review Letters}, vol.~74, no.~11, pp. 1970--1973, 1995.

\bibitem{LiShujun:ReturnMapAttack:IJBC2006}
S.~Li, G.~Chen, and G.~Alvarez, ``Return-map cryptanalysis revisited,''
  \emph{Int. J. Bifurcation and Chaos}, vol.~16, no.~5, pp. 1557--1568, 2006.

\bibitem{LG:ModdingSymmetries:JPIIF96}
C.~Letellier and G.~Gouesbet, ``Topological characterization of reconstructed
  attractors modding out symmetries,'' \emph{J. Phys. II France}, vol.~6,
  no.~11, pp. 1615--1638, 1996.

\bibitem{AlvarezLi:LorenzFrequency2004}
G.~Alvarez, S.~Li, J.~L\"{u}, and G.~Chen, ``Inherent frequency and spatial
  decomposition of the {Lorenz} chaotic attractor,'' arXiv:nlin/0406031, v2,
  2004.

\bibitem{Yang:SpectralCryptanalysis:PLA98}
T.~Yang, L.-B. Yang, and C.-M. Yang, ``Breaking chaotic secure communications
  using a spectogram,'' \emph{Physics Letters A}, vol. 247, no. 1-2, pp.
  105--111, 1998.

\bibitem{AlvarezLi:BreakingPS:CSF2005}
G.~Alvarez, S.~Li, F.~Montoya, M.~Romera, and G.~Pastor, ``Breaking projective
  chaos synchronization secure communication using filtering and generalized
  synchronization,'' \emph{Chaos, Solitons and Fractals}, vol.~24, no.~3, pp.
  775--783, 2005.

\bibitem{Alvarez:BreakingCPM:CSF2004}
G.~Alvarez, F.~Montoya, M.~Romera, and G.~Pastor, ``Breaking parameter
  modulated chaotic secure communication system,'' \emph{Chaos, Solitons and
  Fractals}, vol.~21, no.~4, pp. 783--787, 2004.

\bibitem{Yang:GSCryptanalysis:IEEETCASI97}
T.~Yang, L.-B. Yang, and C.-M. Yang, ``Breaking chaotic switching using
  generalized synchronization: Examples,'' \emph{IEEE Trans. Circuits and
  Systems I}, vol.~45, no.~10, pp. 1062--1067, 1998.

\bibitem{Yang:STZCR:IJCTA95}
T.~Yang, ``Recovery of digital signals from chaotic switching,'' \emph{Int. J.
  Circuit Theory and Applications}, vol.~23, no.~6, pp. 611--615, 1995.

\bibitem{AlvarezLi:ShortTimePeriod}
G.~Alvarez and S.~Li, ``Estimating short-time period to break different types
  of chaotic modulation based secure communications,'' arXiv:nlin.CD/0406039,
  2004.

\bibitem{Storm:DetectingSwitchingEvent:PRE2002}
C.~Storm and W.~J. Freeman, ``Detection and classification of nonlinear dynamic
  switching events,'' \emph{Physical Review E}, vol.~66, no.~5, p. 057202,
  2002.

\bibitem{Short:UnmaskingHyperchaos:PRE98}
K.~M. Short and A.~T. Parker, ``Unmasking a hyperchaotic communication
  scheme,'' \emph{Physical Review E}, vol.~58, no.~1, pp. 1159--1162, 1998.

\bibitem{ZhouLai:ChaoticCryptanalysis:PRE99b}
C.~Zhou and C.-H. Lai, ``Extracting messages masked by chaotic signals of
  time-delay systems,'' \emph{Physical Review E}, vol.~60, no.~1, pp. 320--323,
  1999.

\bibitem{Huang:UnmaskingChaosWavlet:IJBC2001}
X.~Huang, J.~Xu, W.~Huang, and Z.~Lu, ``Unmasking chaotic mask by a wavelet
  multiscale decomposition algorithm,'' \emph{Int. J. Bifurcation and Chaos},
  vol.~11, no.~2, pp. 561--569, 2001.

\bibitem{Alvarez:BreakingPS:Chaos2004}
G.~Alvarez, F.~Montoya, M.~Romera, and G.~Pastor, ``Breaking a secure
  communication scheme based on the phase synchronization of chaotic systems,''
  \emph{Chaos}, vol.~14, no.~2, pp. 274--278, 2004.

\bibitem{Yang:3CC:IEEETCASI97}
T.~Yang, C.~W. Wu, and L.~O. Chua, ``Cryptography based on chaotic systems,''
  \emph{IEEE Trans. Circuits and Systems I}, vol.~44, no.~5, pp. 469--472,
  1997.

\bibitem{Short:ChaoticCrptanalysis:IEEETCASI2001}
A.~T. Parker and K.~M. Short, ``Reconstructing the keystream from a chaotic
  encryption scheme,'' \emph{IEEE Trans. Circuits and Systems I}, vol.~48,
  no.~5, pp. 624--630, 2001.

\bibitem{Murali:3CC:PhysicaD2000}
K.~Murali, ``Heterogeneous chaotic systems based cryptography,'' \emph{Physics
  Letters A}, vol. 272, no.~3, pp. 184--192, 2000.

\bibitem{Murali:CHCS:PRE2001}
------, ``Digital signal transmission with cascaded heterogeneous chaotic
  systems,'' \emph{Physical Review E}, vol.~63, no.~1, p. art. no. 016217,
  2001.

\bibitem{Jiang:2Channels:IEEETCASI2002}
Z.-P. Jiang, ``A note on chaotic secure communication systems,'' \emph{IEEE
  Trans. Circuits and Systems I}, vol.~49, no.~1, pp. 92--96, 2002.

\bibitem{AlvarezLi:Breaking2Channels2006}
A.~Orue, G.~Alvarez, M.~Romera, G.~Pastor, F.~Montoya, and S.~Li, ``Lorenz
  system parameter determination and application to break the security of
  two-channel chaotic cryptosystems,'' arXiv:nlin/0606029, 2006.

\bibitem{BuWang:CSF2004}
S.~Bu and B.-H. Wang, ``Improving the security of chaotic encryption by using a
  simple modulating method,'' \emph{Chaos, Solitons and Fractals}, vol.~19,
  no.~4, pp. 919--924, 2004.

\bibitem{CheeXuBishop:CSF2004}
C.~Y. Chee, D.~Xu, and S.~R. Bishop, ``A zero-crossing approach to uncover the
  mask by chaotic encryption with periodic modulation,'' \emph{Chaos, Solitons
  and Fractals}, vol.~21, no.~5, pp. 1129--1134, 2004.

\bibitem{WuHuZhang:CSF2004}
X.~Wu, H.~Hu, and B.~Zhang, ``Analyzing and improving a chaotic encryption
  method,'' \emph{Chaos, Solitons and Fractals}, vol.~22, no.~2, pp. 367--373,
  2004.

\bibitem{Alvarez:BreakingCPM:CSF2005}
G.~Alvarez, F.~Montoya, M.~Romera, and G.~Pastor, ``Cryptanalyzing an improved
  security modulated chaotic encryption scheme using ciphertext absolute
  value,'' \emph{Chaos, Solitons and Fractals}, vol.~23, no.~5, pp. 1749--1756,
  2005.

\bibitem{Indian:MSCPM:IJBC2001}
P.~Palaniyandi and M.~Lakshmanan, ``Secure digital signal transmission by
  multistep parameter modulation and alternative driving of transmitter
  variables,'' \emph{Int. J. Bifurcation and Chaos}, vol.~11, no.~7, pp.
  2031--2036, 2001.

\bibitem{XuChee:CSKwFS:IJBC2004}
D.~Xu and C.~Y. Chee, ``Chaotic encryption with transient dynamics induced by
  pseudo-random switching keys,'' \emph{Int. J. Bifurcation and Chaos},
  vol.~14, no.~10, pp. 3625--3631, 2004.

\end{thebibliography}

\end{document}